\documentclass[osajnl,twocolumn,showpacs,superscriptaddress,10pt]{revtex4-1} 

\pdfoutput=1

\usepackage{amsmath,amssymb,graphicx}
\usepackage[colorlinks=true, allcolors=blue]{hyperref}

\begin{document}

\title{Compact and versatile laser system for polarization-sensitive stimulated Raman spectroscopy}

\author{Hugo Kerdoncuff}\email{Corresponding author: hk@dfm.dk}
\affiliation{Danish Fundamental Metrology, Matematiktorvet 307, DK-2800 Kgs. Lyngby, Denmark}

\author{Mark R. Pollard}
\affiliation{Danish Fundamental Metrology, Matematiktorvet 307, DK-2800 Kgs. Lyngby, Denmark}

\author{Philip G. Westergaard}
\affiliation{Danish Fundamental Metrology, Matematiktorvet 307, DK-2800 Kgs. Lyngby, Denmark}
\affiliation{Current address: OFS Fitel Denmark ApS, Priorparken 680, DK-2605 Br\o ndby, Denmark}

\author{Jan C. Petersen}
\affiliation{Danish Fundamental Metrology, Matematiktorvet 307, DK-2800 Kgs. Lyngby, Denmark}

\author{Mikael Lassen}
\affiliation{Danish Fundamental Metrology, Matematiktorvet 307, DK-2800 Kgs. Lyngby, Denmark}



\begin{abstract}
We demonstrate a compact and versatile laser system for stimulated Raman spectroscopy (SRS). The system is based on a tunable continuous wave (CW) probe laser combined with a home-built semi-monolithic nanosecond pulsed pump Nd:YVO$_4$ laser at 1064 nm. The CW operation of the probe laser offers narrow linewidth, low noise and the advantage that temporal synchronization with the pump is not required. The laser system enables polarization-sensitive stimulated Raman spectroscopy (PS-SRS) with fast high resolution measurement of the depolarization ratio by simultaneous detection of Raman scattered light in orthogonal polarizations, thus providing information about the symmetry of the Raman-active vibrational modes. Measurements of the depolarization ratios of the carbon-hydrogen (CH) stretching modes in two different polymer samples in the spectral range of 2825--3025~cm$^{-1}$ were performed. Raman spectra are obtained at a sweep rate of 20~nm/s (84~cm$^{-1}$/s) with a resolution of 0.65~cm$^{-1}$. A normalization method is introduced for the direct comparison of the simultaneously acquired orthogonal polarized Raman spectra.
\end{abstract}


\maketitle

\section{Introduction}

There is a growing demand for novel compact laser systems for vibrational spectroscopy and microscopy tools, with the aim of drastically reducing size and cost while increasing reliability and sensitivity. One particular technique, Raman spectroscopy is a label-free method which relies on the inelastic scattering of light due to molecular vibrations and can reveal detailed information about the chemical composition of the sample investigated. Raman spectroscopy has applications in physical, chemical, biological sciences and in a number of industrial areas \cite{Das2011, Lewis2001, Nafie2015}. In the past three decades the use of Raman spectroscopy has greatly increased due to large improvements in laser, detector and filter technologies, as well as in the development of more elaborate detection schemes. Some of these schemes are based on coherent Raman scattering, where specific molecular vibrations are coherently driven by two lasers whose energy differences match the vibrational transition energies. As a result light scattering is enhanced and directed into the spatial modes of the lasers \cite{Potma2012}, thereby facilitating higher collection rates of the Raman scattered light and increasing the detected Raman signal by several orders of magnitude compared to schemes based on spontaneous Raman scattering. Coherent Raman scattering enables very fast and sensitive measurements of Raman spectra, which is particularly relevant for the growing field of hyperspectral imaging \cite{Zhang2015}.

Coherent Raman techniques can be divided into two categories, coherent anti-Stokes Raman scattering (CARS) \cite{Hiramatsu2012, Ideguchi2013, CampJr2014, Yampolsky2014, Cleff2016} and stimulated Raman scattering (SRS) \cite{Freudiger2008, Saar2010, Freudiger2011, Ozeki2012, Karpf2015, Westergaard2015}, where Raman signals are detected from the inelastic scattering of light to higher energies (CARS) or to lower energies (SRS).
CARS signals contain a nonresonant background arising from nonresonant contributions to the nonlinear optical response, which alters the Raman spectra. This background is absent from SRS signals, thus SRS provides an advantage over CARS. Many of the laser systems for CARS and SRS are expensive and bulky laser systems and often based on femtosecond pulses \cite{Ploetz2007, CampJr2014}. Pulses from the lasers require temporal synchronization \cite{Cleff2016, Freudiger2008, Ozeki2012}, thus making the CARS and SRS techniques both complex and cumbersome to operate.

We present a compact and simple laser system based on a tunable continuous wave narrow linewidth probe laser combined with a semi-monolithic nanosecond pulsed  Nd:YVO$_4$ laser at 1064 nm as a pump laser.  The laser system is used for demonstrating SRS and PS-SRS, where we detect the Stokes scattering of an optical pulse from a pump beam to a probe beam. Because the probe beam is CW no temporal synchronization is required which simplifies the operation of the laser system. We have implemented a detection and normalization scheme which allows for simultaneous measurement and comparison of spectra from orthogonal polarizations of the Stokes light. By observing the evolution of the polarization of inelastically scattered light, one may retrieve the third order nonlinear susceptibility tensor, which expresses the vibrational symmetry properties of the sample \cite{Potma2012, Duboisset2015}. Polarization-sensitive Raman spectroscopy has been used for assigning and characterizing vibrational modes via the measurement of their depolarization ratio \cite{Saito2000} or optical activity \cite{Barron1971, Hug1999, Barron2010, Nafie2011, Hiramatsu2012}. Initially performed by spontaneous Raman scattering, polarization-sensitive measurements have recently been carried out with coherent Raman scattering, leading to a significant improvement in speed and sensitivity \cite{Saito2000, Hiramatsu2012, Munhoz2012, Duboisset2015, Cleff2016}.
Information about the molecular structure is very valuable in, for example, synthetic chemistry or polymorph analysis. Polarized Raman spectroscopy is therefore commonly used for the study of macromolecular orientation in crystal lattices or polymer samples. Furthermore, it is non-destructive and suitable for remote analysis unlike other common techniques such as X-ray diffraction or solid-state NMR \cite{Tanaka2006}. We demonstrate the capabilities of our compact laser system by measuring the spectra of two different polymers, polydimethylsiloxane (PDMS) and polymethylmethacrylate (PMMA), in the CH stretching region (2825--3025~cm$^{-1}$). By changing the probe laser we can measure any spectral region making our laser system highly versatile.

\section{Experimental setup}

\begin{figure}[htbp]
\centering
\includegraphics[width=1\linewidth]{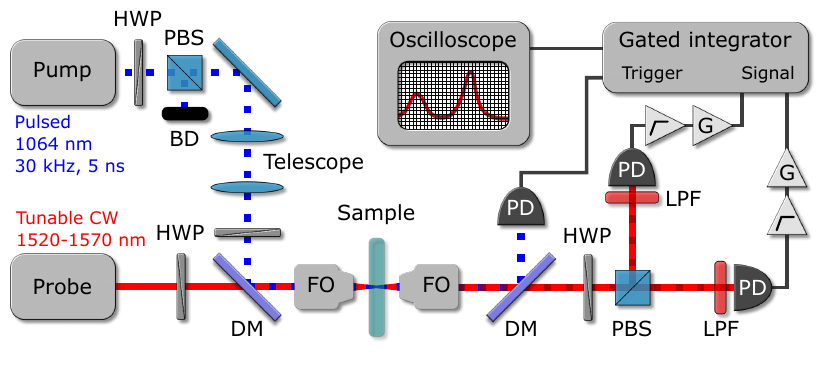}
\caption{SRS setup (see main text for details). BD: Beam dump. PBS: Polarizing beamsplitter. HWP: Half-wave plate. DM: Dichroic mirror. FO: Focusing objective. PD: InGaAS PIN photodetector. LPF: Longpass filter.}
\label{fig:SRG setup}
\end{figure}

Figure \ref{fig:SRG setup} shows the schematics of our SRS setup. The pump beam is generated by pumping a Nd:YVO$_4$ semi-monolithic cavity with a 808~nm broad-area laser diode, resulting in 5 ns-long optical pulses at 1064~nm with a repetition rate of 30 kHz. The pulses are generated due to passive Q- switching using an intra-cavity saturable absorber (SESAM). A combination of half-wave plate (HWP) and polarizing beamsplitter (PBS) enables the adjustment of the pump power as well as the preparation of the pump beam in a known polarization state. The narrow band and low noise probe beam is provided by a widely-tunable CW external-cavity diode laser (New Focus TLB-6728) which allows mode-hop-free tuning of the probe wavelength from 1520~nm to 1570~nm.
The probe beam is filtered through a polarization-maintaining single mode fiber to provide a TEM$_{00}$ mode. The pump laser presents a good beam quality in a TEM$_{00}$ mode allowing an efficient overlap with the probe beam.
The polarizations of the pump and probe beams are adjusted with HWPs before the beams are combined on a dichroic mirror (DM). Due to the CW operation no temporal synchronization of the pump and probe beams is required. The combined beams are directed to a high power focusing objective ($\times$10, 0.25 NA) and overlapped at the focal point in the sample. A telescope is used to adjust the pump beam diameter for spatial mode matching with the probe beam. A second objective ($\times$10, 0.30 NA) collects the light after the sample, which contains the pump and probe light transmitted through the sample, as well as the coherent Raman scattered light. A DM separates the pump beam from the probe beam. A PBS splits the latter into orthogonal polarized components which are each directed onto 150 MHz bandwidth photodetectors (Thorlabs PDA10CF-EC), where additional longpass filters (Thorlabs FEL1100 and Semrock LP02-1064RE) remove any residual pump radiation.

When the frequency difference between the two lasers matches that of an allowed Raman transition, the pump laser will be depleted, while the probe laser will experience a gain in intensity (stimulated Raman gain (SRG)). The SRG signal is obtained by gated integration of the AC signal from each photodetector. We use a bias-tee and a RF amplifier after the photodetectors to extract the AC signal. The pump pulses are measured after the sample and used for triggering the gated integration. Raman spectra are acquired by scanning the probe wavelength via the laser controller and recording the integrated signals on an oscilloscope. The probe wavelength is recorded in the form of a voltage ramp from the laser controller.

\section{Experimental results}

\begin{figure}[htbp]
\centering
\includegraphics[width=1\linewidth]{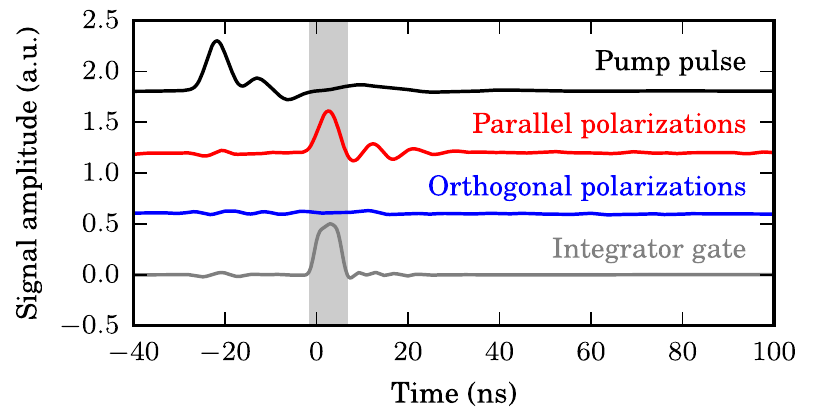}
\caption{Time traces of the detected pump pulse intensity (black), Raman scattering intensity in the parallel (red) and orthogonal (blue) polarization components, and integrator gate (gray), for a PDMS sheet sample probed at 1541~nm. The integration window is highlighted by the light gray strip. Ripples on the time traces are due to limited bandwidth of the detectors and improper impedance matching in the detection. The relatively long time delay between the pump pulse signal and the Raman scattering signals is due to differences in their respective electronic path lengths before acquisition. Traces are offset vertically by steps of 0.6 units for clarity.}
\label{fig:SRG time trace}
\end{figure}

Figure \ref{fig:SRG time trace} shows typical time traces acquired from the detected pump pulse, from the Raman scattering gain in the polarization components which are parallel and orthogonal to the pump beam, and from the integrator gate. Here, a PDMS sheet was probed at 1541~nm which corresponds approximately to the resonance of the symmetric CH stretch in the methyl groups \cite{Smith1984, Jayes2003}. From the time traces it may be deduced that the Raman band of the symmetric CH stretch is strongly polarized because of the large amplitude difference between the Raman scattering signal in the parallel and orthogonal polarizations.

\begin{figure}[htbp]
\centering
\includegraphics[width=1\linewidth]{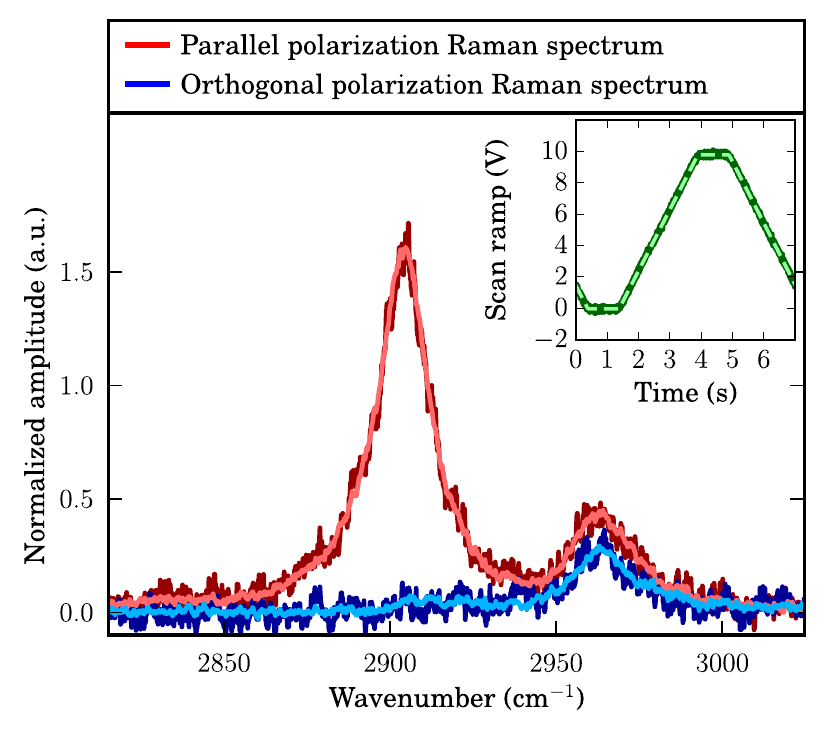}
\caption{Fast-acquisition (2.5 s, one-shot) SRG spectra of the symmetric and antisymmetric CH stretches in a PDMS sample for parallel (red) and orthogonal (blue) polarizations of the pump and probe beams. Dark and light traces indicate one-shot and 10 times averaged spectra, respectively. The normalized spectra result from our normalization procedure (see main text). The inset shows the ramp voltage provided by the laser controller (solid line) which is used for probe wavelength calibration by fitting a trapezoidal waveform (dashed line).}
\label{fig:PDMS fast spectra}
\end{figure}

The spectra of a PDMS sample shown in Fig. \ref{fig:PDMS fast spectra} are recorded by sweeping the probe wavelength from 1520~nm to 1570~nm at a speed of 20.00~nm/s, which is the maximum scanning speed available with our tunable laser source. The gated integrator triggers on every two detected pump pulses at 15 kHz and averages over 100 samples, providing a 150 Hz acquisition rate which corresponds to one data point every 0.15~nm. The resolution of the spectrum is therefore limited by this instrument to 0.65~cm$^{-1}$. With 100~mW average pump power and 4~mW CW probe power the signal-to-noise ratio (SNR) for the two Raman peaks in the parallel polarization spectrum are 34.8 and 9.4 for a one-shot measurement (dark red trace in Fig. \ref{fig:PDMS fast spectra}). Further averaging of successively acquired spectra enhances the SNR (light red trace).

In order to compare the Raman spectra obtained in the orthogonal and parallel polarization configurations we normalize the signal amplitudes to a preliminary reference scan. We modulate the probe beam intensity by driving the laser diode current with pulses of 20 ns duration and 30 kHz repetition rate from a function generator. The gated integrator is triggered by the electrical pulse signal such that it integrates the probe beam intensity modulation measured by the photodetectors. By scanning the probe wavelength, in the absence of the pump beam, we record reference spectra for the two detection and acquisition paths (the complete transfer functions of the system, including the sample), which reflect variations of the probe optical power with wavelength, the transmission of optical components, and etalon effects. Since both the acquired SRG spectra and the reference spectra are proportional to the optical probe intensity, we can remove optical intensity variations by normalizing the SRG spectra to the reference spectra. In the same manner we also normalize to the transfer function of the detection and acquisition. Note that for the symmetric vibrational modes we can further improve the SNR by employing differential measurement detection of the probe laser \cite{Westergaard2015} to reduce the measurement noise to the shot noise level.

\begin{figure}[htbp]
\centering
\includegraphics[width=1\linewidth]{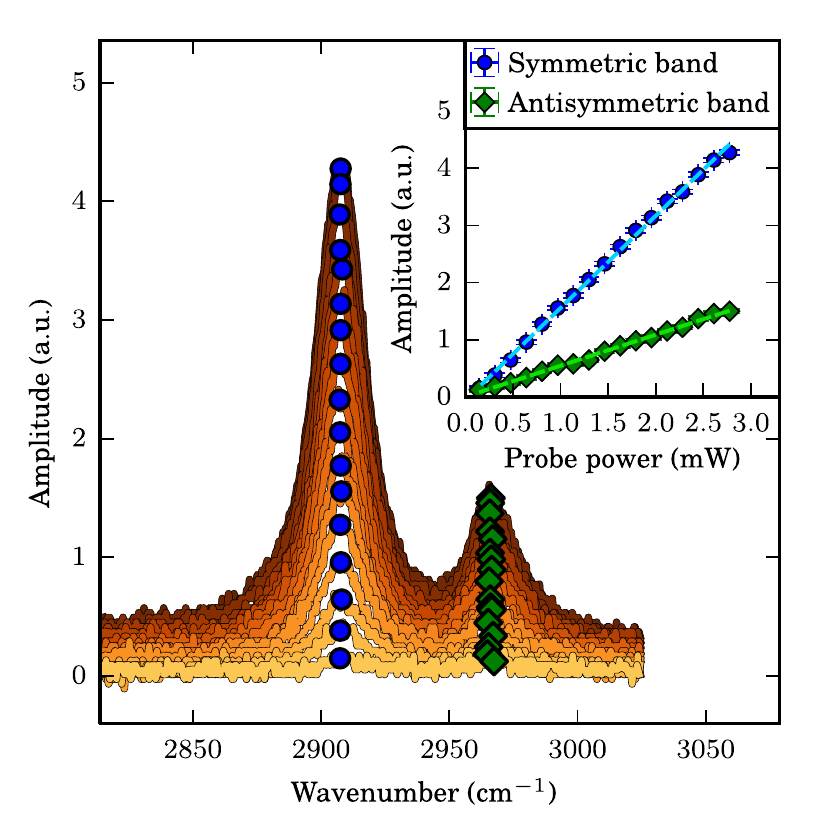}
\caption{SRG spectra at various probe power showing the symmetric and antisymmetric CH stretches in PDMS. The blue dots and green diamonds indicate the peak amplitudes of the SRG signal on resonance with the symmetric and antisymmetric CH stretches, respectively. The inset shows the peak amplitudes for the two Raman bands as a function of probe power. Two linear fits to the data (dashed lines) with correlation coefficients above 0.998 demonstrate the linearity of the SRG signal with probe power. The vertical and horizontal error bars on the data are given by the standard deviations of the spectra noise and of the probe power measurements, respectively.}
\label{fig:linear probe power}
\end{figure}

We verified the linear dependence of the SRG with probe power, as shown in Fig. \ref{fig:linear probe power}. The scanning speed was lowered to 1.20 nm/s and the number of averaged samples was increased to 10000 for a threefold increase of the SNR. The pump power was kept fixed at an average of 200.8~mW (1.34 kW peak power).
The pump and probe powers were measured in front of the sample before each acquisition of the Raman spectrum. The probe power was measured at 1541 nm.
The polarization of the pump and probe are aligned and the spectra are recorded on a single detector. 
Peak amplitudes and wavenumbers of the two observed Raman bands of PDMS were extracted by a peak detection algorithm \cite{Duarte2015}. Linear regressions to the peak amplitudes against probe powers confirm the linearity of the SRG process with the probe power (see inset in Fig.~\ref{fig:linear probe power}).

\begin{figure*}[htbp]
\centering
\includegraphics[width=0.9\linewidth]{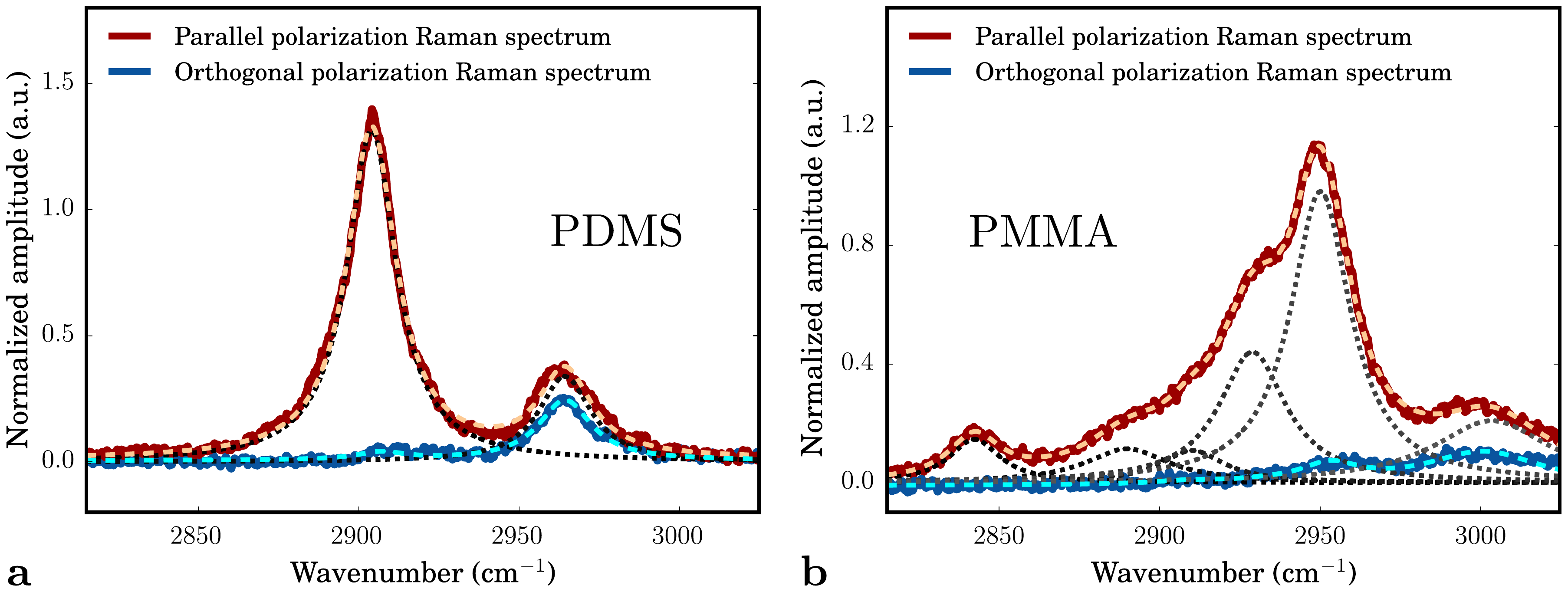}
\caption{Spectra of PDMS (a) and PMMA (b) in the parallel (red) and orthogonal (blue) polarization configurations resulting from averaging 10 fast-acquisition spectra. Fits to the data (dashed lines) use a combination of Lorentzian curves (gray dotted lines) corresponding to the molecule vibration modes.}
\label{fig:fast PMMA spectrum fit}
\end{figure*}

We applied our PS-SRS scheme to two polymer samples, PDMS and PMMA, which present strong Raman bands in the 2825--3025~cm$^{-1}$ region due to methyl groups. The polarization resolved spectra of the two polymers can be described by the theory of Munhoz et al. \cite{Munhoz2012}. 
Figure~\ref{fig:fast PMMA spectrum fit} shows 
the spectra of the two polymers in the parallel and orthogonal polarization configurations. The average pump power was set to 100~mW and the CW probe power to 4.0~mW at 1541~nm. We fit the spectra with a combination of Lorentzian curves 
corresponding to the molecular vibrations present in the polymers \cite{Potma2012}. We select the number of Lorentzian curves for the fit according to the number of peaks and shoulders which can be distinguished on the spectra, and which are reported in the literature \cite{Smith1984, Jayes2003, Willis1969, Dirlikov1980}.

The PDMS sample was prepared from Sylgard 184 silicone elastomer with a square geometry of 25 mm length and a thickness of 3 mm. It presents two Raman bands in the scanned region due to the symmetric and the antisymmetric CH stretches, with resonances given from the fit to the parallel polarization spectrum at 2904.0~cm$^{-1}$ and 2964.2~cm$^{-1}$, respectively. This is in very good agreement with earlier results \cite{Smith1984, Jayes2003}. From the fitting to the spectra we extract depolarization ratios of 0.023 ($\pm$~0.010) and 0.707 ($\pm$~0.059) for the symmetric (strongly polarized) and antisymmetric (depolarized) CH stretches, respectively.

The PMMA sample was purchased from Goodfellows Ltd (product number ME303050) with a thickness of 5 mm. Its Raman spectrum displays three distinct peaks in our scan range which can be attributed to CH stretching modes \cite{Willis1969, Dirlikov1980}. We use a combination of four Lorentzian curves to fit the central peak in order to account for three shoulders which are observed on the low frequency side of the resonance and which are also reported in \cite{Dirlikov1980}. In total our fit model comprises six Lorentzian curves which characterize the vibrational modes. Assignment of molecular vibrations and measurement of depolarization ratios for PMMA are summarized in Table \ref{tab:molecular vibrations PMMA}.
The first four vibration bands are combination bands involving stretching modes of methyl groups and appears strongly polarized. Since their contributions to the orthogonal polarization spectrum are buried in the measurement noise, the SNR sets upper bounds on their depolarization ratios.
Contrary to what is stated by Willis et al. in \cite{Willis1969} we find that the fifth Raman band is also strongly polarized as we would expect from symmetric vibrational modes. The sixth Raman band appears to be weakly polarized corresponding to non-symmetric vibrational modes. On the high wavenumber end of the spectrum, the fit deviates slightly from the experimental data due to the probable presence of a Raman band at the limit of our scan range, around 3025~cm$^{-1}$ \cite{Dirlikov1980}, which we do not take into account in our fit model.

\begin{table}[htbp]
\centering
\caption{\bf CH stretch in PMMA measured with PS-SRS}
\begin{tabular}{ccl}
\hline
Wavenumber & Depol. ratio & Assignment \cite{Willis1969, Dirlikov1980}\\
\hline
2842.4 cm$^{-1}$ & $<$0.07 & $\nu$(O-CH$_3$) \\
2889.9 cm$^{-1}$ & $<$0.09 & $\nu$($\alpha$-CH$_3$) \\
2910.2 cm$^{-1}$ & $<$0.09 & $\nu$(O-CH$_3$) \\
2928.9 cm$^{-1}$ & $<$0.02 & $\nu$($\alpha$-CH$_3$), $\nu_s$(CH$_2$) \\
2950.1 cm$^{-1}$ & 0.065 ($\pm$ 0.011) & $\nu_a$(CH$_2$), \\
 & & $\nu_s$(O-CH$_3$), $\nu_s$($\alpha$-CH$_3$) \\
3003.5 cm$^{-1}$ & 0.535 ($\pm$ 0.058) & $\nu_a$(O-CH$_3$), $\nu_a$($\alpha$-CH$_3$) \\
\hline
\multicolumn{3}{l}{\footnotesize $\nu_s$: symmetric stretching mode, $\nu_a$: antisymmetric stretching mode}
\end{tabular}
  \label{tab:molecular vibrations PMMA}
\end{table}

\section{Conclusion}

We have demonstrated a novel compact and versatile laser system with unique capabilities for performing PS-SRS.  We observed molecular symmetry and measured for the first time (to the best of our knowledge) the depolarization ratios associated with the CH stretches in the methyl groups of PDMS and PMMA, in the range 2825--3025~cm$^{-1}$.
For fast characterization of molecular symmetry we have developed a normalization scheme that enables the simultaneous measurement and comparison of SRG signals in orthogonal polarizations of the coherent Stokes light. This normalization relies on the  linearity of the SRG signal with the probe power which we verified experimentally.

The Raman spectra of PMMA and PDMS samples measured with our system exhibit a high SNR and a high resolution (0.65~cm$^{-1}$) for a short acquisition time (2.5 s). The spectral acquisition time of our system is limited by the wavelength scanning speed of the probe laser and can be greatly improved with current laser technology. However, with 150 Hz spectral point acquisition rate, our system can follow continuously the evolution of one vibrational mode during processes such as the straining or curing of a polymer. On the other hand SNR and resolution can be enhanced to the detriment of acquisition time: by increasing sample or spectrum averaging for the former, and by slowing the scanning speed for the latter. Furthermore, the vibrational frequency range accessible by our system is only limited by the wavelength scanning range of our probe laser but similar widely tunable lasers exist for other wavelength ranges.

The laser system presented in this article is not limited to the study of polymer samples and could in principle be used to analyze liquid or biological samples, among others. We believe that PS-SRS is a promising tool for improving the speed, resolution and sensitivity of molecular symmetry measurements.  For future work we believe that the proposed laser system has the potential to be integrated with a microscope for SRS imaging.
Further improvements to the measurement sensitivity for application in SRS microscopy may involve balanced detection, baseline subtraction, optimization of beam overlap by using achromatic focusing objectives, use of higher magnification objective lenses leading to higher intensity at focus, and stabilization of the pump power and repetition rate.

\section*{Funding}

We acknowledge the financial support from the Danish Agency for Science Technology and Innovation.



\end{document}